\documentclass[aps,pra,superscriptaddress,preprint,onecolumn,showpacs,preprintnumbers,
amsmath,amssymb,floatfix,nofootinbib]{revtex4}
\usepackage{amsthm,graphicx}

\newcommand{\om}{\omega}
\newcommand{\Om}{\Omega}

\newcommand{\s}{\sigma}

\newcommand{\la}{\lambda}
\newcommand{\lo}{\lambda_0}

\newcommand{\e}{\epsilon}
\newcommand{\ve}{\varepsilon}

\newcommand{\psigp}{\psi_{GP}}

\def\r {{\bf r}}

\def\a {\alpha}

\def\g {\gamma}

\begin{document}

\title{Attractive Bose-Einstein Condensates in three dimensions under rotation: 
Revisiting the problem of stability of the ground state in harmonic traps}

\author{Marios C. Tsatsos}\email{marios.tsatsos@pci.uni-heidelberg.de} 
\affiliation{Theoretische Chemie, Physikalisch-Chemisches Institut, 
Universit\"at Heidelberg, Im Neuenheimer Feld 229, D-69120 Heidelberg, Germany}

\date{\today}

\begin{abstract}
We study 
harmonically trapped ultracold Bose gases 
with attractive interparticle 
interactions under external rotation in three spatial dimensions
and 
determine the critical value of the attraction strength where the 
gas collapses as a function of the rotation frequency. 
To this end we examine the stationary state in the corotating frame 
with a many-body approach as well as within the Gross-Pitaevskii theory
of systems 
in traps
with
different anisotropies. 
In contrast to recently reported results 
[N. A. Jamaludin, N. G. Parker, and A. M. Martin, Phys. Rev. A \textbf{77}, 051603(R) (2008)], 
we find that the collapse is not postponed in the presence of rotation. 
Unlike repulsive gases, the properties of the attractive system remain practically unchanged 
under rotation in isotropic and slightly anisotropic traps.
\end{abstract}

\pacs{03.75.Hh, 05.30.Jp, 03.65.-w}

\maketitle

Bose-Einstein condensates (BECs) have occupied a central role in the study 
of atomic and quantum physics since their first experimental realization. 
In particular, attractive condensates \cite{Bradley1995}, i.e., systems whose 
bosons attract each other, 
are distinguished due to their peculiar features. 
Namely, in a trapped three-dimensional attractive gas whose number of particles
or the strength of the interparticle 
interaction
exceeds a threshold value, 
the kinetic energy cannot balance the (negative) interaction energy and so 
the gas implodes and collapses \cite{Collapse-supernova,Adhikari2001}. 
However, in the range below the critical interaction strength (or critical particle number), 
there can exist \emph{metastable} states, i.e., 
states that will survive the collapse for some finite time. 
The collapse of attractive BECs and situations where 
it can be hindered have 
been the subject of much interest already two decades ago, see, for instance, 
Refs.~\cite{Sademelo1991,Dalfovo1996}. 
It is furthermore known that the attractive gas, 
once prepared in a vortex configuration, 
will be more stable against collapse \cite{Shi1997,Dalfovo1996,Adhikari2001}. 
More recently, fragmented metastable excited states \cite{cederbaum:040402} 
and ground states with definite 
nonzero angular momentum \cite{Tsatsos2010} 
have also been found to
postpone the collapse.

On the other hand, 
it is known that rotating (stirring) condensates is a way of imprinting 
angular momentum in a gas and nucleating vortices \cite{Abo-Shaeer2001,Madison2001}.
In repulsive gases rotating with a frequency smaller than the trapping frequency there can exist configurations where the 
system is well described by a stationary state with some finite nonzero vorticity. 
Vortices \cite{Madison2000}, 
vortex lattices \cite{Abo-Shaeer2001} 
and highly correlated -- fractional quantum Hall -- states \cite{Gemelke2010}, 
as well as giant-vortices \cite{Engels2003} have all 
been experimentally observed in repulsive gases. 
In sharp contrast, 
the behavior of the attractive system under rotation is quite different \cite{Kavoulakis2004,Ghosh,prl}. 
The question of how rotation would affect the stability and 
collapse of the attractive condensate in harmonic traps 
has recently been addressed \cite{Jamaludin2008}. 
In Ref.~\cite{Jamaludin2008} it has been found at the 
Gross-Pitaevskii (GP) mean-field (MF) level 
that the attractive gas can be stabilized against collapse for rotation 
frequencies 
smaller than the trap frequency.
These findings have motivated us to attack the same problem at 
the many-body (MB) level. 
We show herein that rotating an attractive condensate, 
confined by a harmonic isotropic or slightly anisotropic trap,
with a frequency below the trap frequency,
does not have an impact on the stability as well as 
on the angular momentum of the ground state. 
We then analyze the problem on the GP (MF) level and find as well 
that no stabilization of the ground state occurs.
We stress at this point that it has been previously shown that there is no stabilization of the attractive gas in an isotropic anharmonic trap with a slight anharmonicity for rotation frequencies below the trapping one (see \cite{Kavoulakis2004}).

\subparagraph{The system.}

We consider an attractive BEC of $N$ atoms of mass $m$, 
confined by a generally anisotropic trapping potential 
\begin{equation}
V(\r)=\frac{1}{2} m \om^2\left[(1-\ve) x^2+(1+\ve) y^2+\zeta^2z^2\right] = V_0(x,y,z) - \ve V_a(x,y),
\label{potential}
\end{equation}
where $\om$, $\ve$ and $\zeta$ are real nonnegative parameters 
that determine the frequencies of the trap and its deformation, 
namely $\om_x=\om \sqrt{1-\ve}, \om_y=\om \sqrt{1+\ve}$ 
and $\om_z=\om\zeta$, $V_0=\frac{m \om^2}{2} (x^2+y^2+\zeta^2 z^2)$ 
is the axially symmetric part of the potential 
and $V_a=\frac{m\om^2}{2} (x^2-y^2)$ the `rotating' anisotropy. 
Since we are interested in the rotating problem we will work in 
the corotating frame of reference, 
where the MB Hamiltonian takes on the time-independent appearance:
\begin{equation}
\hat H = \sum_i^N \left[-\frac{\hbar^2}{2m}\nabla^2(\r_i) + V(\r_i) - 
\Om \hat L_z(\r_i)\right] + \lo\sum_{i<j}^N \delta(\r_i-\r_j),
\label{Hamiltonian}
\end{equation}
where $\Om$ is the frequency of the rotation around 
the z-axis, 
$\hat L_z$ the z-projection of the angular momentum operator and $\lo$ measures 
the interaction strength and takes on negative values for attraction. 
We set hereafter $\hbar=m=\omega=1$ so as to work in dimensionless units.

The Hamiltonian of Eq.~(\ref{Hamiltonian}) admits exact solutions in the absence 
of interaction, i.e., when $\lo=0$. 
In the case of isotropic system ($\ve=0,~\zeta=1$) and in the limit 
of fast rotation ($\Om\rightarrow\om$) the energy levels are 
organized into what is known as \emph{Landau Levels}. 
The same holds true for weak interparticle interactions \cite{Wilkin1998,Mueller,Morris2006}. 
Thus, in the fast rotation and weak interaction 
limit the \emph{Lowest Landau Level} (LLL) is particularly 
designated for the description of the ground state of the system.
The orbitals that comprise the (scaled) LLL have 
the form $\psi^{LLL}_k(\r)=N_k r^k e^{-r^2/2\s^2} Y_k^k(\theta,\phi)$, 
$k=0,1,2,\dots$, where $Y_k^k$ is the spherical harmonic with $l=m_l=k$ and $N_k$ 
is the normalization constant. 
The scaling parameter $\sigma$ defines the width of the Gaussian 
part and will be treated variationally, i.e., 
so as to minimize the total energy. 
Of course, if $\lo=0$ then $\s=1$. 
At the resonance, $\Omega_r=\om$, 
all (infinitely many) orbitals of this set become degenerate in energy. 
The above orbitals can also be expressed in Cartesian coordinates as 
appropriate linear combinations of the 
solutions $\phi_i(x,y,z)=\varphi_{n_x}(\om_x,x)\varphi_{n_y}(\om_y,y)\varphi_{n_z}(\om_z,z)$ 
of the three-dimensional 
harmonic oscillator, i.e., 
the scaled Hermite-Gauss functions, 
$\varphi_{n_x}(\om_x,x)=\frac{(\om_x/\s^2\pi)^{1/4}}{\sqrt{2^{n_x} n_x !}} 
H_{n_x}\left(\frac {\sqrt{\om_x}}{\s} x\right) e^{-\om_x x^2/2\s^2}$, 
where $H_n(\dots)$ denotes the Hermite polynomial of degree $n$. 
Namely, for the isotropic case $\om_x=\om_y=\om_z$, we rewrite the orbitals as 
\begin{equation}
\psi^{}_k(\r)=\sum_{n_x+n_y=k} c_{i} \phi_i, 
\end{equation}
with $c_i=\langle \psi^{LLL}_k |\phi_i \rangle$, 
$n_x+n_y=k$, $k=0,1,2,\dots,$ 
and $i=i(n_x,n_y)=\frac{1}{2}[n_x+3 n_y+(n_x+n_y)^2]$ is a function that we employ to uniquely map the pair $\{n_x,n_y\}$ to the single parameter $i$.
Once we depart from the isotropy of the trap the infinite degeneracy, 
now at $\Omega_r=\om_x < \om_y$, 
is \emph{not} lifted \cite{Ring1980} 
and the above LLL states are not solutions of the anisotropic system. 
Since the radial symmetry of the trap is broken the orbitals do not 
possess exact angular symmetries and one cannot express the solutions 
in terms of pure spherical harmonics anymore. 
Instead, one should resort to the orbitals $\psi_k$ expressed as a mixture of functions $\phi_i$. 
The same transformation coefficients $c_i$, that are defined above for the isotropic case, can also be used for generic $\om_x\neq\om_y\neq\om_z$. 
This transformation maps the functions from the Hermite-Gauss representation to that with nonzero (expectation value of) 
orbital angular momentum. 
Of course, for $\ve=0$ and $\zeta=1$ (i.e., for isotropic traps)
the mapped orbitals give back the spherical harmonics. 
The expectation values of $\hat L_z$ for the orbitals $\psi_k$, 
for $\ve$ small enough, are $\langle \psi_k |\hat L_z| 
\psi_k\rangle \equiv l_k = \left[1+\frac{\ve^2}{8} +\mathcal O(\ve^4) \right]k$, 
with $k=0,1,2,\dots$.
Note that, 
when $\ve\neq 0$, 
the orbital set $\{\psi_k\}$ is also not an exact solution of the 
noninteracting anisotropic Hamiltonian, 
since the linear combination $\psi_k=\sum c_i \phi_i$ mixes nondegenerate states. 
However, in the limit of small $\ve$, 
this choice is justified on account of working with single-particle 
states $\psi_k$ that have nonzero (expectation value of)
angular momentum $l_k$ 
and thus 
allows for a possible coupling to the rotation.

\subparagraph{Many-body approach.}

We study our system at the MB level, i.e., 
beyond a MF description. 
To this end we follow 
the \emph{Configuration Interaction} (CI) expansion, 
a general variational MB method that allows the system 
to fragment and takes into consideration fluctuations of the states. 
For details on this method and the construction of the configuration 
space the reader is referred to the literature, e.g., 
\cite{ModernChem,Haugset,CIE,Tsatsos2010}. 
The MB wave function $|\Psi\rangle$ of the system is expanded over 
a set of functions $|\Phi_i\rangle$ (permanents),
\begin{equation}
|\Psi\rangle = \sum C_i |\Phi_i\rangle,
\end{equation}
each describing a MF state of 
a condensed or fragmented Bose gas of $N$ atoms. 
The permanents are built over a certain set of $M$ single-particle functions (orbitals). 
In this work $M=4$ and the set of orbitals comprise 
the LLL and its anisotropic extension, as described above.
The permanents can be written in an occupation-number-representation as
$|\Phi_i\rangle=|\vec n\rangle=|n_0,n_1,\dots n_{M-1}\rangle,$
where it is meant that $n_i$ bosons occupy the $\phi_i$ orbital, satisfying $\sum_i n_i=N$.
The Hamiltonian of the problem is then represented 
as a matrix $\mathcal H$ over the permanents $|\Phi_i\rangle$ 
and diagonalized. 
The eigenvalues $E_i$ of 
$\mathcal H$ are the energies of the states. 
The eigenvectors $\{C_i\}$ of $\mathcal H$ provide 
us 
with the 
wave functions with which one can compute
various quantities like the 
\emph{natural occupation numbers} $\rho_i$ of the ground and excited states, 
with $\sum_i \rho_i=N$. 
Note that the \emph{natural orbitals} and the orbitals 
described and used in the expansion above coincide.
This holds for the isotropic case
(due to the symmetry of the problem) and has been found (numerically) 
to be well satisfied 
for the slightly anisotropic case
discussed below. From the natural occupations we can calculate the total angular momentum 
of the ground state as $L=\sum_{l=0}^3 l\rho_l$. 
By varying the parameter $\s$ (i.e., the Gaussian width of the orbitals), 
we minimize the energies per particle $\e=E/N$ 
as a function of the rotation frequency $\Om$, 
for some fixed value of $\la=|\lo|(N-1)$ 
and determine 
the optimal value $\s_0$. 
The analysis of the system that follows is always done 
for optimal states, i.e., at $\s_0$. 
The number of particles is hereafter set to $N=12$.

\begin{figure}
\centering
\includegraphics[scale=0.8]{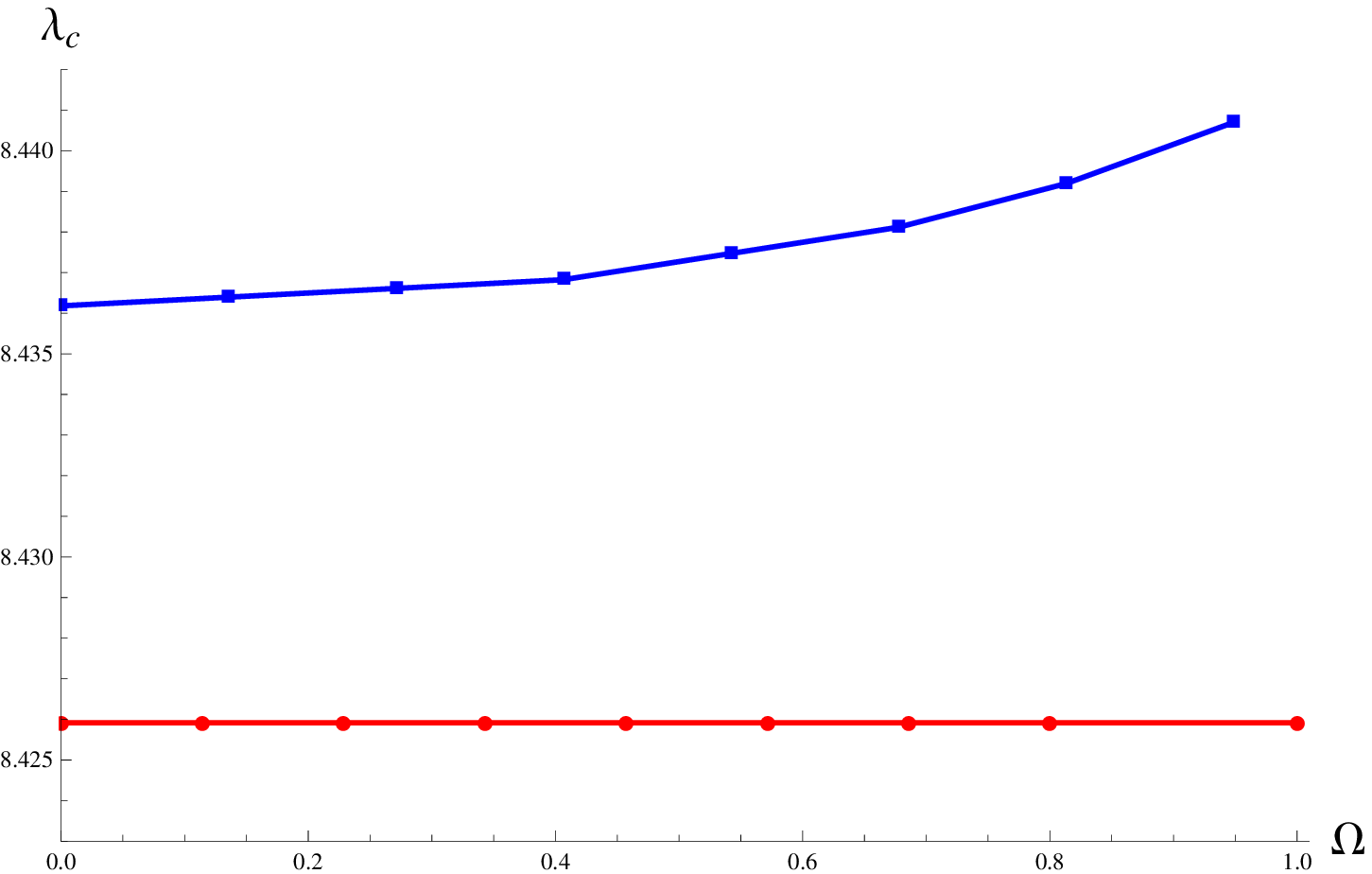}
\caption{(Color online). Many-body calculations for the 
critical parameter $\la_{c}$ as a function of the 
frequency of the external rotation $\Om$, 
for the cases of isotropic [$\ve=0$, $\zeta=1$; lower (red) line] 
and slightly anisotropic [$\ve=0.1$, $\zeta=1$; upper (blue) line]
confining traps. 
The critical interaction $\la_{c}$ remains 
practically unaffected (note the scale!) for the whole region of 
$\Om\in[0,\Omega_r=\sqrt{1-\ve})$. 
The number of particles is $N=12$. 
See text for more details.
All quantities are dimensionless.\\ \\}
\label{Fig-cr-om}
\end{figure}

We denote with $\la_c$ the critical value of the parameter 
$\la$ where the ground state of the condensate ceases to exist. 
This is calculated as the largest value of $\la$ where there 
is a (local) minimum in the 
energy $E$ as a function of $\s$. 
The absence of such a minimum denotes a collapsed state 
(see also \cite{Fetter,Pethick_Smith,cederbaum:040402,Tsatsos2010}). 
We are interested in the dependence of $\la_c$ on the rotation frequency $\Om$. 
In Fig.~\ref{Fig-cr-om} we plot the critical value $\la_c$ against $\Om$ 
for the isotropic $\ve=0,~\zeta=1$ and the slightly anisotropic 
case $\ve=0.1,~\zeta=1$.~\footnote{Note that even smaller trap anisotropies are sufficient to 
nucleate vortices in experimental setups, 
like $\ve=0.025$ for instance, in the rotating repulsive 
gas of Ref.~\cite{Madison2001}.} 
The values of $\Om$ range from 
$0$ to $\Omega_r=\omega_x=\sqrt{1-\ve}$. 
At exactly the 
resonance 
frequency $\Omega_r$, 
the energy diverges and the gas becomes mechanically unstable. 
We notice no change in the stability of the ground state of 
the isotropic system as the rotation frequency $\Om$ increases 
from $0$ up to $\Omega_r$, 
and only a negligible increase in $\la_c(\Om)$ of less than $0.1\%$ for $\ve=0.1$. 
The value of the critical parameter $\la_c=8.425(9)$ remains unchanged when $\ve=0$ 
for the whole allowed region of $\Om$, 
and marginally increases from 
$\la_c(0)=8.436(2)$ to 
$\la_c(\Omega_r)=8.440(7)$ for $\ve=0.1$.~\footnote{Here and hereafter,
when we write $\la_c(\Omega_r)$ it is meant, 
mathematically,
$\la_c(\Omega)$ in the limit 
of the resonance 
frequency $\Omega \to \Omega_r$. 
The same is meant for other system's properties
at the resonance frequency.}   
These results, 
obtained at the MB level, 
obviously contradict the GP results of Ref.~\cite{Jamaludin2008} 
(see analysis and discussion below).

Next, to analyze the MB results, 
we chose $\la=3$ as a representative value of the interaction 
parameter of an isotropic system ($\ve=0$) with noncollapsed 
ground state and calculated the energy per particle $\e$, 
the angular momentum per particle $L/N$ and 
the natural occupations $\rho_i$, $i=0,\dots,3$ for the ground state. 
We found that the above quantities remain 
constant for any $\Om \in [0,\Omega_r)$. 
The state remains condensed ($\rho_0=N$), 
carries no angular momentum ($L/N=0$) 
and has energy $\e=1.396(8)$. 
For the anisotropic case of $\ve=0.1$ we 
also found that the above quantities practically do not change.
Namely,
$\rho_0$ marginally decreases from $\rho_0(0)\simeq 12$ to $\rho_0(\Omega_r)=11.998(9)$, 
and the rest of the natural occupations 
change 
from $\rho_1(0) \simeq 10^{-8}, \rho_2(0)\simeq 10^{-5}, \rho_3(0)\simeq 10^{-13}$ to 
$\rho_1(\Omega_r)\simeq 10^{-7}, \rho_2(\Omega_r)=10^{-3}, \rho_3(\Omega_r)\simeq 10^{-12}$.
There is an insignificant decrease in the energy 
[from $\e(0)=1.395(8)$ to $\e(\Omega_r)=1.395(4)$] 
and a 
corresponding increase in the angular momentum 
[from $L(0)/N\simeq 0$ to $L(\Omega_r)/N=2 \cdot 10^{-4}$].

The fact that the ground state of the isotropic system is 
found to be fully (i.e., 100\%) condensed deserves some discussion. 
The total absence of depletion and fluctuations in this case is explained 
if one considers the MB 
orbital set used: 
since each orbital $\psi^{LLL}_k$ has 
different angular symmetry a coupling between the different modes is forbidden 
due to the symmetry of the problem. 
Any nonzero occupation of 
the $i=1,2,3$ orbitals would 
result in the change of the total angular momentum of the system. 
Naturally, such a coupling is induced in the system when 
the anisotropy $\ve$ is turned on and hence the
occupations $\rho_i,i=1,2,3$ can be nonzero.
Nonetheless,
as we have found above,
for attractive systems in three-dimensional isotropic and
slightly anisotropic traps,
coupling of the ground zero-angular-momentum state
to excited-states with nonvanishing angular momentum
essentially does not occur,
even for rotation frequencies as high as the resonance 
frequency $\Omega_r$.
In other words,
for rotating attractive BECs 
none of the $\psi^{LLL}_{k>0}$ 
(or, 
for slightly anisotropic traps, $\psi_{k>0}$)    
states becomes the state lowest-in-energy,
even for rotation frequencies as high as the resonance 
frequency $\Omega_r$.

Do 
the above findings 
change for a MB basis set that does allow for ground-state depletion? 
The answer is negative.
Having used, in place of the LLL, the set consisting of the $s,p_+,p_0$ and $p_-$ orbitals (see in this 
respect Ref.~\cite{Tsatsos2010}), we found the ground state of the isotropic system slightly depleted 
(i.e., about 98\% condensed for $\lambda \simeq \lambda_c$),
but its angular momentum zero for all rotation frequencies up to the resonance 
frequency $\Omega_r$.
Side by side,
the depletion and fluctuations of the ground state
do not depend on the rotation frequency.
Importantly,
the critical value $\lambda_c$ for the collapse 
does not depend on the rotation frequency as well.
The same conclusion holds
for slightly anisotropic traps ($\varepsilon=0.1$).
In summary,
we have shown by a MB approach that the critical value
of the interaction for collapse, $\lambda_c$, 
of rotating three-dimensional attractive BECs
does not depend on the frequency of rotation.

The fact that the ground states of both the isotropic ($\ve=0$) 
and the slightly anisotropic system ($\ve=0.1$) were found at the MB level 
to be essentially fully condensed for any rotation frequency $\Om$
smaller than the resonance frequency $\Omega_r$, 
means that the GP theory should be valid here and 
reproduce the MB conclusions.

\subparagraph{Analysis within the Gross-Pitaevskii approach.}

We now want to turn from the MB to the GP 
(MF) description and 
address the same question, 
namely how the stability of the attractive gas 
is affected as the system is rotated externally.
The GP theory assumes that all particles reside 
in the same single-particle state and hence the 
wave function for the state of the whole 
system is given by a single permanent
$\Psi_{GP}=\prod_i^N \psigp(\r_i)$.
The GP orbital $\psigp$ for the ground state of the 
rotating gas should be represented with an 
ansatz that takes 
into consideration orbitals with nonzero angular momentum, 
as done in the MB treatment. To this end we expand $\psigp$ as a linear combination
\begin{equation}
\psigp(\r,\s)=\sum_k b_k\psi_k(\r,\s),
\label{GPansatz}
\end{equation}
where the basis $\psi_k$ is the same as the one used in 
the MB computations reported above. The coefficients $b_k$ and 
parameter $\s$ (Gaussian width of the orbitals) are determined variationally 
with the normalization constraint $\sum_k |b_k|^2=1$ and 
the summation running over from $k=0$ to $k=3$. 
We calculate the expectation value $E=\langle \Psi_{GP}|\hat H|\Psi_{GP}\rangle$ 
with the above GP 
ansatz and minimize it with respect to 
the parameters $b_i, i=1,2,3$ and $\s$, for different 
values of the interaction parameter $\la=|\lo| (N-1)$ 
and for given values of $\Om \in [0,\sqrt{1-\ve})$ 
and the (small) trap anisotropy $\ve$. 
The expectation value of the angular momentum 
operator for $\psigp$ is $l= \sum_{i,j} \mathcal L_{ij} b^{\ast}_i b_j $,
$i,j=1,\ldots,4$, 
where the matrix elements $\mathcal L_{ij}=\langle \psi_i | \hat L_z | 
\psi_j \rangle$ are given, in second order approximation, 
as $\mathcal L_{ii}=l_{i-1}=\left(1+\frac{\ve^2}{8}\right)\cdot(i-1)$, 
$\mathcal L_{13}= \mathcal L_{31} = \frac{\ve}{2\sqrt{2}}$, 
$\mathcal L_{24}=\mathcal L_{42}=\frac{\sqrt 3 \ve}{2\sqrt 2}$ 
and the rest of the elements are zero. The total angular momentum is $L=N l$.

We calculate the critical value of the interaction $\la_c$ 
as a function of the rotation frequency $\Om$, 
for the cases of $\ve=0,~\zeta=1$ and $\ve=0.1,~\zeta=1$. 
As anticipated from the MB analysis, 
we again found no essential change in the stability of the gas, 
as $\Om$ varies from $0$ to $\Omega_r=\sqrt{1-\ve}$. 
Namely, in the isotropic case, 
the GP ansatz 
of Eq.~(\ref{GPansatz}) yields 
the value $\la_c=8.425(9)$ which coincides with that 
obtained from the MB analysis, 
and remains fixed 
for any $\Om\in[0,\Omega_r)$. 
In the case of anisotropic trap ($\ve=0.1$) 
we found $\la_c(0)=8.436(3)$ 
and a negligible increase 
as $\Om$ increases, i.e., 
$\la_c(\Omega_r) \simeq 1.0005\cdot\la_c(0)$.
We then fix the interaction parameter to $\la=3$ as before. 
In the isotropic case, $\ve=0$, the energy $\e=1.396(8)$ 
and angular momentum $L/N=0$ remain constant 
for all $\Om\in[0,\Omega_r)$
and,
as above,
the values coincide with those of the MB ansatz.
For $\ve=0.1$,
we found $\e(0)=1.395(8)$, 
$\e(\Omega_r)=1.391(8)$, 
$L(0)/N\simeq0$ and $L(\Omega_r)/N=0.045(2)$.
Namely, 
the energies found in the MB and GP 
approaches are almost identical 
while the angular momentum computed
within the GP theory at the resonance frequency $\Omega_r$ 
is somewhat 
above the value that the MB theory gives. 
Nonetheless, both values of angular momentum 
can be considered practically zero.
In conclusion, the rotation does not increase the stability of the ground 
state described by the GP 
ansatz of Eq.~(\ref{GPansatz}).

Last, we re-examine the attractive rotating gas using a different GP ansatz that has been previously used in the literature, namely the ansatz of Ref.~\cite{Jamaludin2008} (see also references therein). 
The authors of Ref.~\cite{Jamaludin2008} 
considered a GP ansatz for 
the ground state of the system, 
which they expressed -- depending on the geometry of the confining potential -- either 
as a Gaussian-sech single-particle wave function:
\begin{equation}
\phi(\r)=\left[N(2l_xl_yl_z \pi)^{-1}\right]^{1/2} e^{-x^2/2{l_x}^2} 
e^{-y^2/2{l_y}^2} sech\left(\frac{z}{l_z}\right)e^{i\alpha xy}
\label{orbital1}
\end{equation}
or as a Gaussian:
\begin{equation}
\phi(\r)=\left[N (l_xl_yl_z)^{-1} \pi^{-3/2}\right]^{1/2} 
e^{-x^2/2{l_x}^2} e^{-y^2/2{l_y}^2} e^{-z^2/2l_z^2} e^{i\alpha xy}.
\label{orbital2}
\end{equation}
The parameters $l_x, l_y, l_z$ and $\a$ were to be determined variationally. 
The phase $\a xy$ put in Eqs.~(\ref{orbital1})-(\ref{orbital2}) is referred to as 
the `quadrupolar flow' term; a nonzero 
value of $\alpha$ increases the energy 
of the isotropic system in this state.

In Ref.~\cite{Jamaludin2008} it is found that, 
using the ansatz of Eq.~(\ref{orbital1}) or (\ref{orbital2}), 
the stability of the gas is significantly increased with increasing frequency $\Om$. 
However, an algebraic error in the above 
work is responsible 
for this (erroneous) behavior of the energy per particle $\e$ as a function 
of the rotation 
frequency $\Om$. 
Already in Eq.~(4) of Ref.~\cite{Jamaludin2008} there is a sign error; 
redoing carefully the calculations we convinced 
ourselves that in the 
true expression the sign in the last term of the integrand is a plus instead of a minus. 
This sign error gives rise to an extra (negative) term in the GP energy functional 
which overestimates the dependence of the energy on $\Om$ and artificially 
reduces the energy of the system 
(see the Appendix for more details).

In fact, also with the 
ans\"atze of Eqs.~(\ref{orbital1})-(\ref{orbital2}) 
the external rotation does not practically affect the stability of the condensate, 
in the sense that the critical value of the interaction parameter $\la_c$ 
does not essentially change with $\Om$. 
We have verified this by calculating the energies and critical parameters $\la_c$ 
by varying all three parameters $l_x,l_y,l_z$ of Eqs.~(\ref{orbital1})-(\ref{orbital2}) 
for $\zeta=0,1,5$ and $\ve=0,0.1$, as it is originally done 
in Ref.~\cite{Jamaludin2008} 
(the parameter $\alpha$ is expressed as a function of $l_x$ and $l_y$
and absorbed into the GP energy functional as in \cite{Jamaludin2008}). 
The Gaussian-sech ansatz of Eq.~(\ref{orbital1}) is used in the case $\zeta=0$, 
while the Gaussian ansatz of Eq.~(\ref{orbital2}) is used when $\zeta=1 $ and $5$. 
The critical $\la_c$ of the radially symmetric 
systems 
(i.e., $\varepsilon=0$, $\zeta=0,1,5$)
remains fixed, 
while in the slightly anisotropic systems 
(i.e., $\varepsilon=0.1$, $\zeta=0,1,5$)
$\la_c$ does not 
increase more than $0.2\%$ as $\Om$ increases 
from $0$ to $\Omega_r$. 
The computed 
values of $\la_c$ for different values of $\zeta$ and $\ve$ are presented in Table~\ref{table}. 
We then fix $\la=3$ and calculate the energies and angular momenta of the ground state. 
For $\ve=0$, irrespective of the choice of $\zeta$, 
the energy has been found to be independent of the rotation frequency $\Om$. 
For instance, we found $\e(\ve=0,\zeta=1)=1.396(8)$.
For $\ve=0.1$ the energy $\e(\Omega_r)$ 
was found to decrease by 1.5\% for $\zeta=0$, 
by 0.32\% for $\zeta=1$, and by 0.04\% for $\zeta=5$ 
with respect to the corresponding energy $\e(0)$ of the nonrotating system. 
Last, 
the expectation values 
of the angular momentum of the two 
ans\"atze 
are (almost) exactly 
zero for $\ve=0$ ($\ve=0.1$), 
regardless 
of the value of 
$\zeta$.

Finally, we point out that, for the cases examined, the optimal value of parameter $\alpha$ is practically zero. 
Note that the minimization of the GP energy functional 
with the ansatz of Eq.~(\ref{GPansatz}) yields 
a distribution 
for the coefficients $b_i$ 
($b_0\simeq1$ and $b_i\simeq0,~i=1,2,3$) 
which essentially includes 
only the first of the LLL Gaussian-shaped orbital. 
I.e.,
the ans\"atze of Eqs.~(\ref{orbital2}) and (\ref{GPansatz}) 
essentially 
coincide with the respective 
orbital of 
isotropic systems.

\subparagraph{Summary and Outlook.} 

We have studied the stability 
under rotation
of attractive ultracold
Bose gases, 
confined by an isotropic as well as a slightly 
anisotropic harmonic trap.
The problem has been mapped to and 
calculated in the corotating frame.
Both many-body and Gross-Pitaevskii approaches revealed that
the rotation does not affect the stability of the gas against the collapse. 
Namely, the maximum value of the interaction strength $\la_c$ 
where the attractive gas collapses remains essentially 
unchanged as the rotation frequency $\Om$ varies within 
the extreme values $0$ and $\Omega_r=\om_x=\sqrt{1-\ve}$, 
where $\om_x$ is the frequency of the 
trap in the direction of the weakest confinement.

We have found here on 
both the MB and the GP (MF) levels 
that the ground state of the rotating attractive system 
carries zero (or almost zero in the anisotropic case) angular momentum
for the whole range of the allowed values of $\Om$.
Obviously, 
no vortex states are created.
In the MB treatment,
this means that 
no transition between LLL states of 
different angular symmetries has been found for the rotating attractive system.
In the GP (MF) 
analysis this means that 
no symmetry broken states were found to 
be energetically favorable as the rotation frequency $\Om$ increases 
from $0$ to $\Omega_r$. 
In both MB and GP approaches, 
the energy of the ground state remains practically unchanged 
and the attractive gas is
condensed in the nodeless 
s-orbital 
as the rotation frequency increases.
Hence, the GP description agrees 
well with the MB 
computation. 
These results conflict the 
findings of Ref.~\cite{Jamaludin2008}.

We revisited then the problem using the ansatz 
that incorporates a `quadrupolar flow' term, used in Ref.~\cite{Jamaludin2008}. 
The energy and stability of the system were again found 
not to be affected by the rotation of the trap. 
The resolution of this discrepancy lies in a sign error in the expectation 
value of the Hamiltonian of Ref.~\cite{Jamaludin2008}, 
which 
leads to a qualitatively 
different behavior 
of the properties of the system as a function 
of the rotation frequency $\Om$. 
Our results are in agreement with
findings in the literature for isotropic harmonic (see \cite{prl})
and isotropic anharmonic traps with
slight anharmonicity \cite{Kavoulakis2004,Ghosh}. 

We should, finally, 
stress that the present variational 
approach to the stationary ground state is not an extensive 
study of the rotating attractive gas. 
Even though we 
can rule out the stability-enhancement of the 
stationary
ground state and the vortex nucleation in the attractive 
gas rotating with a frequency $\Om$ smaller than the resonance
frequency $\Omega_r$, 
there is more physics beyond that. 
For instance, 
the stability of low-lying excited 
states with nonzero total angular momentum is expected 
to depend on $\Om$.~\footnote{Indeed, we have some numerical indication for 
such a dependence in a MB treatment of the problem involving ground states of $L>0$.} 
We have shown elsewhere \cite{Tsatsos2010} that 
ground states with $L>0$ are generally fragmented states 
and thus more stable against collapse. 
A ground state with $L=0$ will not couple 
to the external rotation and hence the ground-state symmetry will not change as $\Om$ increases. 
On the other hand,
a ground state initially with nonzero $L$ 
\emph{will be} affected by 
the rotation. 
The critical parameter $\la_c$ in that case could increase as a function of $\Om$, 
before the latter reaches the extreme value $\Omega_r$, 
and this will further stabilize
the rotating $L \ne 0$ state 
against collapse. 
It is still to be investigated whether and for which parameters' 
values crossings of energy levels and symmetry changing 
of the (noncollapsed) rotating ground-state might occur. 
Last, 
based on the found absence of symmetry breaking 
of the ground state
in the examined region $\Om<\Omega_r$ and the divergence 
of the energy and angular momentum 
for $\Om \ge \Omega_r$, 
we may speculate that, 
in the rotating attractive gas a vortex ground state -- if at all can exist -- 
may only appear as a giant-vortex 
(i.e., a single vortex at the center of the trap whose radius and
vorticity are increasing functions of time) 
and this only for rotation with frequency 
$\Om \ge \Omega_r$. 
A time-dependent many-body treatment,
for instance using 
the multiconfigurational time-dependent 
Hartree for bosons (MCTDHB) method \cite{MCTDHB} that
has described successfully many-body dynamics 
of attractive BECs \cite{MCTDHB_attractive},
should shed light on this interesting 
problem and uncover the response mechanisms 
of attractive gases to rotations.

\begin{acknowledgments}

I am indebted to O. E. Alon, A. I. Streltsov and L. S. Cederbaum for providing useful and essential comments on the manuscript. 
Financial support by the HGSFP/LGFG is acknowledged.

\end{acknowledgments}

\appendix*
\section{The Gross-Pitaevskii energy functional with the `Quadrupolar Flow' Ansatz \label{appendix}}

We re-derive and discuss 
the expression for the energy functional of Ref.~\cite{Jamaludin2008}.
The GP energy functional in the co-rotating frame reads:
\begin{equation}
E=\int \left[\frac{\hbar^2}{2m} |\nabla \phi|^2+V(\r)|\phi|^2+\frac{\lo}{2}
|\phi|^4 + i\hbar \Om (\phi^\ast x \frac{\partial \phi}{\partial y} + 
\phi y \frac{\partial \phi^\ast}{\partial x})\right]d\r,
\label{EnergyJam}
\end{equation}
where $\lo$ measures the strength of the interaction, 
$m$ is the mass of the particle and $\Om$ 
is the frequency of the rotation around the z-axis. 
In Ref.~\cite{Jamaludin2008} there is an algebraic 
error in the above expression (Eq.~(4) of Ref.~\cite{Jamaludin2008}). 
There, the sign of the last term of the integrand is 
a minus instead of a plus. 
This sign error remains further in the calculations 
of Ref.~\cite{Jamaludin2008} and is seen in Eq.~(8) and (10) therein. 
Indeed, in the last term of Eq.~(10) of Ref.~\cite{Jamaludin2008} 
the `-2' term has to be omitted and so the corrected expression would read:
\begin{equation}
\e_G=\frac{1}{4}\left[\frac{1}{\g_x^2}+\frac{1}{\g_y^2}+\frac{1}{\g_z^2} + 
(1-\ve)\g_x^2 + (1+\ve)\g_y^2 + \zeta^2 \g_z^2 \right] - 
\frac{k}{\sqrt{2\pi} \g_x \g_y \g_z} - \frac{\Om^2}{4} 
\frac{\left(\g_x^2-\g_y^2\right)^2}{\g_x^2+\g_y^2}
\label{enjam2}
\end{equation}
(for the $\alpha>0$ branch),
where $\g_{x,y}=l_{x,y}/\sqrt{\hbar/m\om}$ and $k=|\lo| N/4\pi$. 
The same correction is required for Eq.~(8) 
of Ref.~\cite{Jamaludin2008} as well. 
The presence of this extra term 
gives rise to an artificial dependence 
of the critical value of $\lambda_{c}$ on
the rotation frequency $\Om$, 
qualitatively different from 
the correct one. 
Indeed, 
a first order expansion of the (correct) 
energy, Eq.~(\ref{enjam2}), 
around $\gamma_x=\gamma_y$, i.e., 
for small deformations, 
will result in an expression of the energy that 
does not depend on the frequency $\Om$. 
According to this, 
for zero or small ellipticity $\ve$ 
of the trapping potential, 
the resulting 
shape of the orbital $\phi$ is symmetric around the z-axis, 
i.e., $\g_x=\g_y$, 
and the energy of the system, 
as well as the critical interaction strength, 
practically do not depend on the frequency $\Om$. 
On the other hand, the (incorrect) energy $\e_G$ 
as it is calculated 
in Ref.~\cite{Jamaludin2008} strongly depends on $\Om$.
Furthermore, it can be 
easily seen that the `quadrupolar flow'
ansatz of either Eq.~(\ref{orbital1}) or (\ref{orbital2}) 
gives, for small $\ve$, 
an expectation value of almost zero angular momentum 
$\langle \hat L_z \rangle=\frac{1}{2}\frac{(l_x^2 - l_y^2)^2}{l_x^2 + l_y^2} 
m N \Omega \stackrel{\ve\rightarrow 0}{=} 0$, 
and hence cannot describe any state with nonzero angular momentum 
that can in principal increase the stability of the system. 
For zero or small $\ve$ the energy and the critical 
parameter $\la_c$ cannot change as a function of $\Om$ 
and this reflects the cylindrical symmetry of the ans\"atze used, 
since $l_x\simeq l_y$ if $\ve\simeq0$.

% Bibliography

% figures
\newpage

\begin{table}
\centering
\begin{tabular}{| l | | c | c | c | }
\hline
\multicolumn{4}{|c|}{$\la_c$} \\
\hline
& $\zeta$=0 & $\zeta$=1 & $\zeta$=5 \\ \hline
   $\ve$=0    &    9.547(7)  & 8.425(9) & 5.522(8) \\ \hline
   $\ve$=0.1  &    9.554(9)  & 8.429(0) & 5.523(2) \\
\hline
\end{tabular}
\caption{Critical parameter $\la_c$ for 
different values of the anisotropy $\ve$ 
and the z-deformation $\zeta$ of the trapping potential 
calculated in the GP theory, 
using the `quadrupolar flow' ans\"atze, 
see Eqs.~(\ref{orbital1})-(\ref{orbital2}) and text below it. 
The parameter $\la_c$ in the 
radially symmetric 
($\varepsilon=0$) cases does 
not depend on the frequency $\Om$ of the rotation, 
while the change in $\la_c$ as $\Om$ varies from $0$ 
(for which $\la_c$ values are collected in the table) 
to $\Omega_r=\sqrt{1-\ve}$ 
is negligible (less than 0.2\%) when the trap 
is slightly anisotropic ($\ve=0.1$).
All quantities are dimensionless.}
\label{table}
\end{table}

\end{document}